\documentclass[ traditabstract]{aa}
\usepackage{graphicx}
\usepackage{txfonts}
\usepackage{bm}
\usepackage[colorlinks, linkcolor=blue, anchorcolor=blue, citecolor=blue]{hyperref}
\def\etal{{et al.}}
\def\kms{{\rm km\ s$^{-1}$}}

\begin{document}
\title{Chromospheric Magnetic Reconnection caused by Photospheric Flux Emergence: \
	Implications for Jet-like Events Formation}

\author{J. Y. Ding\inst{1,2}
\and M. S. Madjarska\inst{1}
\and J. G. Doyle\inst{1}
\and Q. M. Lu\inst{2}}

\institute{
Armagh Observatory, College Hill, Armagh BT61 9DG, N. Ireland \\
\email{jyd@arm.ac.uk}
\and School of Earth and Space Sciences, University of Science and Technology 
of China, Hefei 230026, China }

\authorrunning{J. Y. Ding et al.}
\titlerunning{Chromospheric Magnetic Reconnection caused by Photospheric Flux Emergence}

\abstract{
Magnetic reconnection in the low atmosphere, e.g. chromosphere, is investigated 
in various physical environments. Its implications for the origination of explosive 
events (small--scale jets) are discussed. 
A 2.5-dimensional resistive magnetohydrodynamic (MHD) model in Cartesian coordinates
is used. It is found that the temperature and velocity of the outflow 
jets as a result of magnetic reconnection are strongly dependent on the physical 
environments, e.g. the magnitude of the magnetic field strength and the plasma 
density. If the magnetic field strength is weak and the density is high, the 
temperature of the jets is very low ($\sim$$10^4$~K) as well as its velocity 
($\sim$$40$~{\rm km~s$^{-1}$}). However, if environments with stronger magnetic 
field strength (20~G) and smaller density (electron density $N_e=2\times 10^{10}$~cm$^{-3}$)
are considered, the outflow jets reach  higher 
temperatures of up to $6\times 10^5$~K and a line-of-sight velocity of up to 
$130$~{\rm km~s$^{-1}$} which is comparable with the observational values of 
jet-like events.} 

\maketitle

\section{Introduction} \label{sec::intro}

Jet-like events, first reported by Brueckner \& Bartoe (\cite{brueck83}), are
characterised by non-Gaussian spectral line profiles. Dere \etal\ (\cite{dere91})
suggested that they are produced by bi-directional jets as a result of 
magnetic reconnection. To date, these jet-like events (often called explosive events) 
are mainly observed in  spectral lines
formed at transition region temperatures (Dere \cite{dere94}; Chae \etal\ \cite{chae98}; Innes
\etal\ \cite{innes01}; Madjarska \& Doyle \cite{mad03}), although observations of explosive
events in chromospheric lines are also reported. For example, Madjarska \& Doyle
(\cite{mad02}) presented the temporal evolution of different temperature plasma using
high cadence (10~$\rm s$) observations obtained with the Solar Ultraviolet 
Measurement of Emitted Radiation (SUMER) spectrometer, and found a
time delay in the response of the S~{\sc vi}~933~\AA~($2\times 10^5$~K) line with
respect to Ly~6 ($2\times 10^4$~K), with the Ly~6 line responding
earlier. They concluded that the jet-like events may first appear at
chromospheric temperatures. In follow-up work, Doyle \etal\ (\cite{doyle05}) reported on
a joint SUMER, Coronal Diagnostic Spectrometer (CDS) on board the Solar Heliospheric 
Observatory and TRACE imager study, confirming the possibility that some
jet-like events originate in the chromosphere. They further suggested that
jet-like events could be divided into two types: one formed in the
chromosphere and the other formed in the transition region. Some of the 
observed features are the result of spicules and/or macrospicules (Madjarska \& Doyle 
\cite{mad03}; Madjarska \etal\ \cite{mad06}), while others are the result of high velocity 
flows in small loops (Teriaca \etal\ \cite{teria04}). {In more recent work,
Madjarska \etal\ (\cite{mad09}) presented observational data relating explosive events to a surge 
and demonstrated that the division of small-scale transient events into a
number of different subgroups, for instance explosive events, blinkers,
spicules, surges or just brightenings, is ambiguous, implying that the
definition of a feature based only on either spectroscopic or imaging characteristics
as well as insufficient spectral and spatial resolution can be incomplete.}

Several numerical models were developed to study jet-like events.
Sarro \etal\ (\cite{sarro99}) used a 1D magnetic flux-tube model to
simulate the temporal evolution of UV emission line profiles, e.g. C~{\sc iv}~1548.2~\AA,
in response to energy perturbations located below the transition-region.
The maximum blue-shifts they obtained reach values of the order of 100 \kms.
Innes \& T\'oth (\cite{innes99}) 
presented a 2D MHD study on jet-like events with different
initial conditions, representative of different regions in which the
reconnection occurs, e.g. the corona and chromosphere. Their conclusion 
was that high-velocity components in the profiles of lines formed around 
$10^5$ K can be obtained in both cases, irrespective of the initial 
conditions. However, heat conduction was not included,
and no brightening was found at the zero velocity position 
of the spectral line.
In their model, the initial equilibrium state 
consists of two regions of oppositely directed magnetic field lines, with  
a narrow current sheet between the two regions. Magnetic 
reconnection at the current sheet is initiated by introducing localized 
anomalous resistivity.
Roussev \etal\ (\cite{roussev01a}) carried out 2D MHD simulations, where
jet-like events are formed during the process of magnetic reconnection.
In their model, the initial magnetic field is parallel to the $y-$axis (vertical),
and there is a thin current concentration formed along the $y$-axis.
Magnetic reconnection is initiated by a localized increase of the magnetic diffusivity 
in the current concentration.
Blue-shifts of the order of 100 \kms{} were obtained.
By using the same model, they further extended the work and performed simulations
under different physical conditions (Roussev \etal\ \cite{roussev01b}, \cite{roussev01c}).

Yokoyama \& Shibata (\cite{yokoy95, yokoy96}) performed 2D magnetic reconnection 
to study coronal X-ray jets using both oblique and vertical initial coronal magnetic fields. 
The temperature of the hot X-ray jets they obtained reach 3 times the coronal temperature.
Moreno-Insertis \etal\ (\cite{more08}) considered
magnetic reconnection triggered by flux emergence from below the photosphere using a 3D MHD model.
Very strong X-ray jets with high temperature ($3\times 10^7$~K 
at the reconnection site) and high velocity (peak velocity 400~\kms) were produced.
In their model, the flux emergence is very strong (maximum field strength 3.8~kG),
which plays a very important role in forming such strong jets.
Isobe \etal\ (\cite{isobe08}) focused on the process of magnetic flux emergence, and 
presented simulations of magnetic flux emergence driven by 
the upward convective motion.
They found that small-scale horizontal magnetic fields could be produced even when the initial magnetic
field is uniform and vertical.
The horizontal magnetic fields emerging from the convection zone into the photosphere
undergoes magnetic reconnection with the background vertical field,
which is a source of high-frequency MHD waves that may contribute to
coronal heating or solar wind acceleration.

Murray \etal\ (\cite{murray09}) presented another simulation of magnetic
flux emergence, where the long-term evolution of magnetic reconnection was initiated by
flux emergence. A series of reconnection reversals (or oscillatory reconnection) was reported.
All the flux emergence studies mentioned above did not include heat conduction and radiative effects.
The latter will reduce both the temperature and the velocity of the outflow jets.

Litvinenko \& Chae (\cite{litvi09}) discussed 
magnetic reconnection at different heights in the solar atmosphere, and found that
the temperature and speed of the outflow jets vary by several orders.
Their study was based on an extended Sweet-Parker model (Parker \cite{parker57, parker63}), 
assuming that the inflowing magnetic energy is completely converted
in the current sheet into the thermal and kinetic energies of the
outflowing plasma.

In the present study, we use a 2.5-dimensional resistive MHD model 
in Cartesian coordinates to investigate  magnetic reconnection in 
the low atmosphere, e.g. chromosphere. We discuss its implications 
for jet-like events.
Here, we use various physical environments 
representing the lower atmosphere with different magnetic field strengths and 
densities. In our model, the magnetic reconnection is triggered by newly 
emerging magnetic fluxes from below the photosphere which reconnect with the
pre-existing background magnetic field lines in the chromosphere.

The physical model is described in Section 2, while the numerical results are shown in
Section 3. No background heating is included in this model, the implications of which are
discussed in Sect.~\ref{sec::dis}.

\section{Physical Model and Numerical Methods}

\subsection{Basic Equations}

For 2.5-dimensional MHD studies in Cartesian coordinates ($x,y,z$), 
one may introduce a magnetic flux function $\psi(t,x,y)$ to express the 
magnetic field by
$$
	{\mathbf B} = \bigtriangledown\times \left (
		{\psi}\hat{z} \right ) + B_z \hat{z}. \eqno (1)
$$ 
The 2.5-D resistive MHD equations are in the following non-dimensional form
$$
	\frac{\partial\rho}{\partial t} + \bigtriangledown\cdot(\rho {\mathbf v})=0, \eqno (2)
$$
$$
	\frac{\partial \mathbf{v}}{\partial t}+{\mathbf v}\cdot{\bigtriangledown}{\mathbf v}
		+\frac{1}{\rho}\bigtriangledown p - \frac{1}{\rho}{\mathbf j} \times {\mathbf B}-{\mathbf g}=0,
			\eqno (3)
$$
$$
	\frac{\partial \psi}{\partial t} + {\mathbf v}\cdot \bigtriangledown \psi-
		\eta\bigtriangleup\psi = 0,  \eqno (4)
$$
$$
	\frac{\partial B_z}{\partial t}+{\mathbf v}\cdot{\bigtriangledown B_z}
		+ B_z\bigtriangledown\cdot{\mathbf v} - {\mathbf B}\cdot{\bigtriangledown v_z}
		- \bigtriangledown\cdot(\eta\bigtriangledown B_z)  = 0, \eqno (5)
$$
$$
	\frac{\partial T}{\partial t}+{\mathbf v}\cdot\bigtriangledown T 
		+ (\gamma-1)T\bigtriangledown\cdot{\mathbf v} - \frac{2(\gamma-1)\eta}{\rho \beta_0}
		{\mathbf j}\cdot{\mathbf j} - \frac{C_1}{\rho}Q 
		+ \frac{C_2}{\rho}L_r= 0 , \eqno (6)
$$
where $\rho$ is the mass density, $\mathbf v$ is the flow velocity, 
$T$ is the temperature, $p=\rho T$ is the gas pressure, $\mathbf g$ is 
the gravitational acceleration,  $\gamma (=5/3)$ is the adiabatic index, 
$\eta$ is the dimensionless magnetic diffusivity which is inverse to the
magnetic Reynolds number, $\beta_0$ is the characteristic ratio of the 
gas pressure to the magnetic pressure, $\mathbf j=\bigtriangledown \times 
{\mathbf B}$ is the electric current density, $Q=\bigtriangledown \cdot 
[T^{5/2}({\mathbf B}\cdot \bigtriangledown T){\mathbf B}/B^2]$ is the 
field-aligned heat conduction function, and $L_r$ is the radiative losses. 
Here we assume that the plasma is fully ionized, therefore, the 
dimensionless form of $L_r$ is explicitly given by
$$
	L_r = \rho^2 \Lambda (T),  \eqno (7)
$$
where $\Lambda (T)$ is the radiative loss function, of which the expression 
used is that given by Rosner \etal\ (\cite{rosner78}). Various physical environments 
representing the solar atmosphere with different characteristic values are 
examined. The different cases can be divided into two groups according to 
the characteristic value of $\beta_0$. The two groups are denoted by 
Exp. A and Exp. B, corresponding to $\beta_0=0.33 \mbox{ and } 0.033$, 
respectively. For both groups, the following characteristic values
are kept invariant and taken as basic units: $\rho_0=3.33 \times 
10^{-10}$ kg m$^{-3}$ for the mass density, $T_0=10^4$ K for the temperature, 
$L_0=500$ km for the length, and $v_0=\sqrt{R T_0}=12.8$ {\rm km s$^{-1}$} 
for the velocity, where R is the gas constant. In terms of the expression 
$\beta_0=2\mu\rho_0 R T_0/B_0^2$ where $\mu$ is the vacuum magnetic 
permeability, we can derive the characteristic values of the magnetic 
field strength $B_0$ = 6 G for Exp. A and 20 G for Exp. B.
The dimensionless coefficient of the heat conduction and radiative 
losses are given by
$$
C_1=(\gamma-1)\kappa_0 T_0^{7/2}/(\rho_0 L_0 v_0^3) \ \mbox{ and } C_2=(\gamma-1)L_0\rho_0/(v_0^3 m_p^2),
$$
respectively, where $\kappa_0=10^{-11}$ W m$^{-1}$ K$^{-7/2}$
is the classical conductivity coefficient, and $m_p$ is the proton mass.
In the present study the $x-$ and $z-$axes are in the horizontal plane, and $y-$axis represents 
the height of the solar atmosphere.  Therefore, the dimensionless form of the gravitational acceleration is
expressed as 
$$
{\mathbf g}=-g\hat{y},
$$
where $g=0.271 \mbox{ km s}^{-2} L_0/v_0^2$.

\subsection{Initial State}

\begin{figure}[htbp]
\begin{center}
\resizebox{\hsize}{!}{\includegraphics{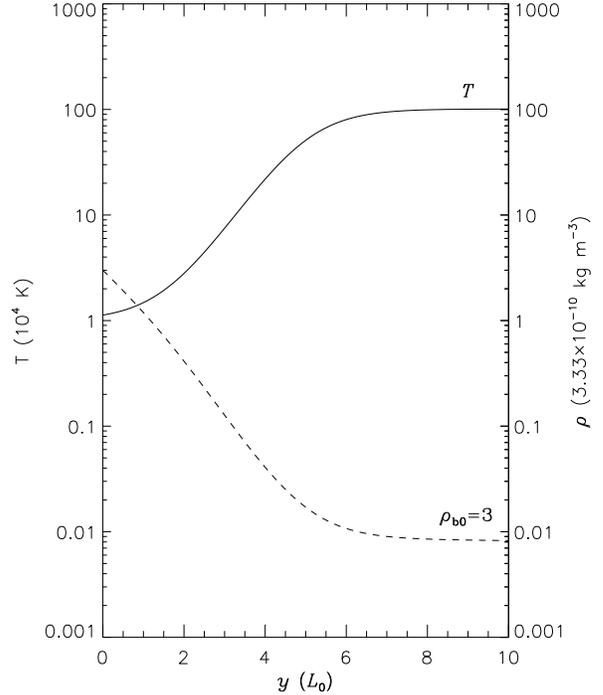}}
\caption{The distribution of the initial temperature ($T$, \emph{solid line}) and 
mass density ($\rho$, \emph{dashed line}) with height ($y$). The density distribution 
corresponds to the case where $\rho_{b0}=3$.}
\label{fig1}
\end{center}
\end{figure}

The initial magnetic field is a linear force-free field taken to be in the 
following form
$$ 
	\psi=\left \{ \begin{array}{ll}
		((4\omega/\pi)\cos(\pi x/2\omega)-4\omega/\pi){\sqrt{2\mu\rho_0 R T_0/\beta_0}}, &	\mbox{ if }|x| < \omega \\
		(-2 |x|-4\omega/\pi+2\omega){\sqrt{2\mu\rho_0 R T_0/\beta_0}},	&	\mbox{ if }|x| \ge \omega,	
		\end{array} \right. \eqno (8) 
$$
$$
	B_z=\left \{ \begin{array}{ll}
		2\cos(\pi x/2 \omega){\sqrt{2\mu\rho_0 R T_0/\beta_0}}, & \mbox{ if }|x| < \omega \\
		0,	&\mbox{ if }|x| \ge \omega,
		\end{array} \right.  \eqno (9)
$$ 
where $\omega$ is the half width of the force-free region (current sheet) 
and is set to  0.1. 
{The unit of magnetic field strength is taken as
$B_0=\sqrt{2\mu\rho_0 R T_0/\beta_0}$. The lower $\beta_0$ is, the stronger the magnetic field strength.}
In order to represent the solar atmosphere from the 
chromosphere to the upper transition region, the profile of the initial 
temperature $T$ is given by
$$
	T(0,x,y)=T_{\mbox{\scriptsize in}}=1+\bigtriangleup T 
		\left\{1+\tanh 
			\left[\epsilon(y-y_s) \right]
		\right\}
$$
where $2\bigtriangleup T=100$ defines the relative temperature jump 
across the transition region, $\epsilon=0.6$ describes the steepness of 
the temperature profile and $y_s=5$ is the position of the middle transition 
region. The temperature is equal to 1 at the bottom and is {101} at the top.
Then, the mass density is calculated from the hydrostatic equilibrium 
equation, $\bigtriangledown p = - \rho g\hat{y}$. In order to obtain a 
unique solution for mass density, we specify the density value at the 
bottom as a boundary condition. It is easy to deduce that both the 
temperature and the mass density of the initial state are dependent 
only on the $y-$coordinate, and are uniform along the $x-$coordinate.
Therefore, the initial mass density at the bottom can be expressed by 
one parameter, $\rho_{b0}$. Fig.~\ref{fig1} shows one example for the initial 
distribution of the temperature and the density along the $y-$coordinate 
when $\rho_{b0}=3$.

In this study, two initial states with different values of $\rho_{b0}$ 
are considered for each of Exp. A and B. In total, four cases which 
represent four physical environments are discussed here, all of which
are listed below:
\begin{itemize}

	\item Exp. A1: $\beta_0=0.33$, $\rho_{b0}=3$ (weak magnetic field and high mass density)

	\item Exp. A2: $\beta_0=0.33$, $\rho_{b0}=0.1$ (weak magnetic field and low mass density)

	\item Exp. B1: $\beta_0=0.033$, $\rho_{b0}=3$ (strong magnetic field and high mass density)

	\item Exp. B2: $\beta_0=0.033$, $\rho_{b0}=0.1$ (strong magnetic field and low mass density)

\end{itemize}
In all the cases, the dimensionless form of the initial magnetic 
field is the same, and the change of the magnetic field strength is only 
due to the change of $\beta_0$. Moreover, the dimensionless form of the 
magnetic flux emergence is the same in all the cases, as well as the 
dimensionless form of the magnetic diffusivity, which will be discussed 
in the following section.

We should note that most numerical studies consider an artificial hearting source, 
however we do not consider it here. The main reason is that we do not know where 
that artificial heating term comes from, thus,
in our model, the initial solar background atmosphere is not in thermal equilibrium.
In order to reduce the influence of the flows that are driven by the re-distribution
of the heat and radiative cooling from the initial background atmosphere, 
the calculations in our paper are carried out over a shorter interval (about 80 s),
compared to the time scale for heat conduction and cooling from the initial configuration.

\subsection{New Magnetic Flux Emergence}

Here, we do not discuss the specific physical mechanism for the 
flux emergence which is beyond the scope of this paper, although 
recent observations with the Hinode Solar Optical
Telescope (SOT; Tsuneta \etal\ \cite{tsuneta08}) reveal that the emergence of
granular-scale magnetic flux is ubiquitous on the solar surface (Centeno 
\etal\ \cite{centeno07}; Ishikawa \etal\ \cite{ishi08}; Isobe \etal\ \cite{isobe08}; 
Okamoto \etal\ \cite{okamo08}). Instead, what we are interested 
in is the physical response of the chromosphere after the magnetic flux 
has emerged from below the photosphere and reached the upper atmosphere, the 
chromosphere and transition region. The emergence of new magnetic flux is implemented numerically by 
changing the boundary conditions of the magnetic field at the bottom, which 
is described below in detail. First, we specify the flux emergence region 
that is taken to be $1 \le x \le 4$. Then, we change the magnetic flux 
function at $y=0$ in the emergence region to a new value, $\psi_e$,
according to a function of $x$ and time, expressed by
$$
	\psi_e=\psi_{\mbox{\scriptsize in}}(x)+
		\alpha|(\psi_{\mbox{\scriptsize in}}(x)-\psi_{\mbox{\scriptsize in}}(1))\times
		(\psi_{\mbox{\scriptsize in}}(x)-\psi_{\mbox{\scriptsize in}}(4))|t/t_e, 
$$
$$
	1 \le x \le 4, \ \ 0 \le t \le t_e, \eqno (10)
$$
where $\psi_{\mbox{\scriptsize in}}$ is the initial magnetic flux function 
at the base, $t_e$ is the duration for the flux emergence, and $\alpha$ 
controls the magnitude and orientation of the emerging flux. When $t$ is 
greater than $t_e$, the newly specified flux distribution at the base 
remains invariant in the emergence region. The flux distribution outside 
the emergence region is fixed to be the same as that of the initial field, 
see equation (8). In all numerical examples discussed below, $t_e$ is 
taken to be $80$ s and $\alpha$ is $-1.2$. 

As new flux emerges, a current sheet will be formed at the interface 
between the newly emerging and pre-existing magnetic fluxes. Then, 
anomalous resistivity is introduced to initiate magnetic reconnection. 
The distribution of the resistivity is localised, and taken to be of  
the following form
$$
	\eta=\left \{\begin{array}{ll}
					\eta_m \mbox{min}(1,j/j_c),  & \ \ \mbox{if }|x| \ge 1, y \le 3, \\
					0, &\ \ \mbox{elsewhere},
					\end{array}
			\right.  \eqno (11)
$$
where $\eta_m=0.002$, and $j_c=10$, is the critical value of the 
current density ($j$). No magnetic reconnection is considered at the 
current sheet of the initial background magnetic field within $|x| 
< \omega$, see equation (11).

There are thus two current sheets in our model. One is at the interface 
between the newly emerging magnetic flux and the pre-existing background
magnetic flux and the other is along $y$ at around $x=0$ which belongs to
the background magnetic field. This is because we only want to study the 
magnetic reconnection occurring on the former one. Here, we introduce 
a localized resistivity (at $|x|>1, y<3$). The current is almost 0 
outside this region ($|x|>1, y<3$), except for that around $x=0$.

The initial state used here contains a current
sheet around $x=0$, which is not the same as the one 
used by Yohokyama \& Shibata (\cite{yokoy95,yokoy96})
(where the initial magnetic field is a uniform vertical one).
The reason for using non-uniform magnetic field  is that 
it is easier for us to compare the results in this paper with those
where reconnections at both current sheets are considered.
The two current sheets represent different magnetic topology: one is
formed between open vertical field lines while 
the other is between open and closed ones. 
The reconnection between the newly emerging and pre-existing
magnetic fluxes will have influence 
on the reconnection between the open vertical fluxes and vice versa.
In this case, the polarity of the emerging flux will be an important factor.
However, in this paper, based on the fact that no magnetic diffusion is
considered at the initial current sheet around $x=0$,
the influence of the initial current sheet on the results 
can be regarded as the influence of different left-boundary conditions
if compared with the uniform vertical field case.
As the reconnection occurs on the right-hand side of the emerging flux arcade,
far away from the initial current sheet around $x=0$,
there is no significant differences in the results if a uniform initial
field is used instead.

\subsection{Computational Domain} \label{sec::grid}

The dimensionless size of the computational domain is $-5\le x \le 5$ 
and $0 \le y \le 10$.
Because of symmetry, we only implement calculation in the right half 
region ($0 \le x \le 5,0 \le y \le 10$).
As for the boundary conditions, we fix all quantities at the base, 
and treat the top ($y=10$) and the right-hand side ($x=5$) as open 
boundaries. Furthermore, we use symmetrical conditions for the left boundary 
($x=0$). A multistep implicit scheme (Hu \cite{hu89}) is used to solve equations (2)-(6).

We adopt $399\times 400$ grid points, which is based on the consideration of the balance between
resolution and computation time. We can also adopt more grid points so as to get better
resolution, but the computation time will become very long then, especially for the case in
which strong magnetic field strength and low mass density are presented. 
Uniform meshes are adopted in both the $x-$ and $y-$ directions.

\section{Numerical Results}

We present here  magnetic reconnection which  occurs in the low
atmosphere and is triggered by newly emerging magnetic flux.
Four cases representing different physical environments are discussed. 
According to the strength of the magnetic field, the results are divided 
into two groups: Exp.~A and B.
For each group, two initial states with different mass density at the 
bottom, $\rho_{b0}$, are considered.

\subsection{Exp. A ($\beta_0=0.33$)}

\begin{figure*}[htbp]
\begin{center}
\resizebox{\hsize}{!}{\includegraphics[width=17cm]{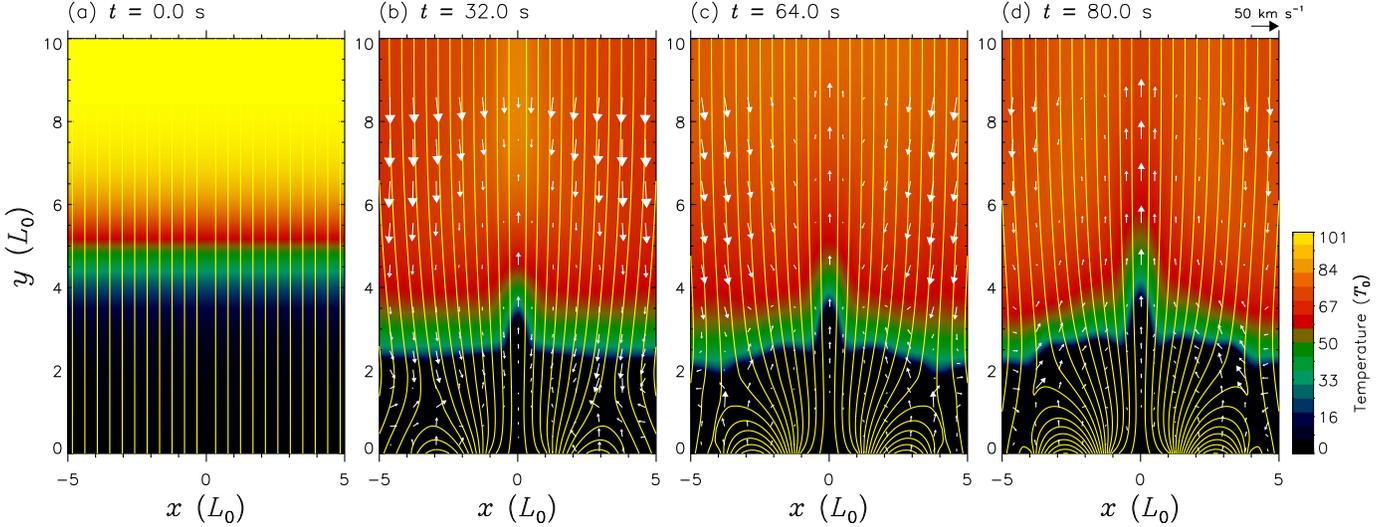}}
\caption{The evolution of the magnetic field (\emph{solid line}), temperature (\emph{colour}) and 
velocity (\emph{arrows}), corresponding to Exp.~A1, where $\beta_0=0.33 \mbox{ and }\rho_{b0}=3$.}
\label{fig2}
\end{center}
\end{figure*}

First, let us discuss the case where $\rho_{b0}=3$ (i.e. electron density
$N_e=6\times 10^{11}$~cm$^{-3}$), which means the mass 
density at the bottom is large, corresponding to Exp.~A1. The results are 
shown in Fig.~\ref{fig2}, where the evolution of the magnetic field, temperature 
and velocity are displayed. 
At the beginning of the experiment, downward flows are 
seen, especially in the middle temperature region (see Fig.~\ref{fig2}b), 
which is caused by heat conduction and radiative cooling. As the new flux emerges, an 
electric current sheet is formed at the right interface between the 
newly emerging and pre-existing magnetic fluxes. Then magnetic 
reconnection occurs at the current sheet producing  bi-directional 
outflows: upward and downward. The outflow jet in each direction will 
be accelerated by magnetic tension forces and shocks along the boundary 
of the jet (Petschek \cite{petsc64}), and reaches its maximum velocity  
($v_{\mbox{\scriptsize jet}}$) somewhere in front of the jet.
For example, at $t=32$ s, the upward flow jet reaches its maximum velocity, 
$v_{\mbox{\scriptsize jet}}=40$ {\rm km s$^{-1}$}, at $x=3.2,y=1.1$, 
whereas the diffusion region is located at a lower altitude, around 
$x=4.0,y=0.3$. As the magnetic reconnection continues, the upward flow 
jet propagating towards higher altitudes will interact with the downward 
flow of the background field, forming a shock at the interface 
between them. The shock will impede the downward flow, so that the 
downward flow first becomes weaker and then even reverses its direction, 
see Fig.~\ref{fig2}. During the process of magnetic reconnection, both the 
magnitude and orientation of $v_{\mbox{\scriptsize jet}}$ change 
dynamically. But, no matter how  $v_{\mbox{\scriptsize jet}}$ changes, 
it is far smaller than that of the jet-like events observed by SUMER.  
In spite of this, it is comparable with the characteristic local 
Alfv\'en speed around the diffusion region, as predicted by the fast 
magnetic reconnection model (Petschek \cite{petsc64}; Priest \& Forbes \cite{priest86}). 
In the following sections,  
the velocity $v_{\mbox{\scriptsize jet}}$ we give is the maximum one 
during the time from $0-80$ s. Also, the velocity in the 
line-of-sight relative to $v_{\mbox{\scriptsize jet}}$ will be 
presented. In this case, the downward flow jet (as a result of magnetic 
reconnection) can not be clearly shown, because the X-point is very 
low in our model. Its height is  0.2 above the bottom at $t=32$ s, and 
is 0.8 at $t=80$ s. The height is so small that only a small part of 
the downward flow jet is displayed in our computation domain. Moreover, 
as mentioned above, the outflow jet will reach its maximum velocity 
somewhere in front of the diffusion region, which means the maximum 
velocity of the downflow jet may be located below the bottom of our 
computational domain. The downward flow jet that can be displayed 
in our model reaches 12 {\rm km s$^{-1}$} at $t=80$~s.
In the following discussions, we will focus on the upflow jet.

\begin{figure}[htbp]
\begin{center}
\resizebox{\hsize}{!}{\includegraphics[width=17cm]{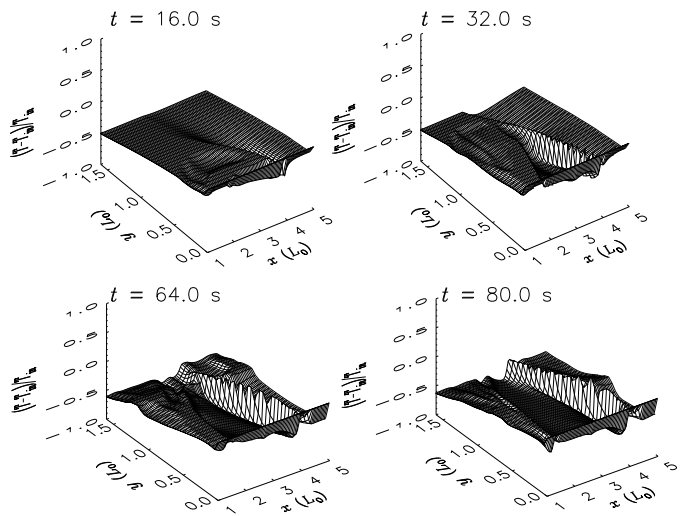}}
\caption{Relative changes in the temperature, 
$(T-T_{\mbox{\scriptsize in}})/T_{\mbox{\scriptsize in}}$ for Exp.~A1, 
with respect to the initial temperature ($T_{\mbox{\scriptsize in}}$),
are shown at four times in the region $1 \le x \le 5, 0 \le y \le 1.6$.}
\label{fig3}
\end{center}
\end{figure}

\begin{figure}[htbp]
\begin{center}
\resizebox{\hsize}{!}{\includegraphics[width=17cm]{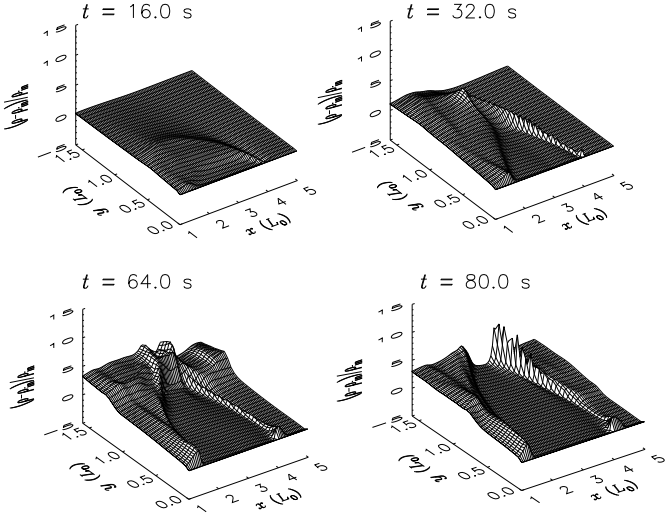}}
\caption{Relative changes in mass density, $(\rho-\rho_{\mbox{\scriptsize in}})/\rho_{\mbox{\scriptsize in}}$
for Exp.~A1, with respect to 
the initial density ($\rho_{\mbox{\scriptsize in}}$), are shown at the same times and in the same region 
as shown in Fig.~\ref{fig3}.}
\label{fig4}
\end{center}
\end{figure}

Fig.~\ref{fig2} is drawn in such a large scale that the temperature evolution 
around the magnetic reconnection region has a very poor resolution. 
In order to better illustrate the temperature evolution in this region, 
we plot only a small part ($1 \le x \le 5, 0 \le y \le 1.6$) of the 
computation domain in Fig.~\ref{fig3} (and in successive plots), where the relative changes in the 
temperature, $(T-T_{\mbox{\scriptsize in}})/T_{\mbox{\scriptsize in}}$, 
with respect to the initial temperature, $T_{\mbox{\scriptsize in}}$,
are shown. The relative 
changes in mass density, 
$(\rho-\rho_{\mbox{\scriptsize in}})/\rho_{\mbox{\scriptsize in}}$, 
with respect to the initial density, $\rho_{\mbox{\scriptsize in}}$, 
are shown in Fig.~\ref{fig4} at the same times and in the same region as in 
Fig.~\ref{fig3}.

Peak structures are clearly seen on each rim of the flux emergence region for both the relative changes in densities and temperatures.
As the magnetic reconnection occurs, the plasma in the diffusion region is heated, and the hot plasma is pulled out by the magnetic tension
force in either direction along the current concentration, resulting in the peak structure on the right-hand side of the flux emergence
region in both Fig.~\ref{fig3} and \ref{fig4}, compared with those in the inflow region. Along the peak structure, the temperature
reaches  its maximum in the diffusion region, while  the  density is  minimum in the diffusion region.
After the magnetic reconnection, part of the reconnected magnetic field lines extend outward towards $x=0$ (see Fig.~\ref{fig2}), resulting in 
a negative horizontal velocity component of the upflow jet.
Accordingly, on the left-hand side of the flux emergence region, 
plasma accumulates and a strong shock is created which causes the formation
of the another peak structure in the relative changes in density plot; 
the peak structure in the relative changes in temperature plot
is due to the heating effect of the shock.
In the following, the temperatures at two locations are analysed: one in the diffusion 
region and the other at the location of $v_{\mbox{\scriptsize jet}}$. 
They are denoted by $T_{\mbox{\scriptsize max}}$ and $T_{\mbox{\scriptsize jet}}$, respectively.

Even though the plasma temperature along the current concentration is heated through magnetic reconnection, 
the relative changes in temperature are negative all over the region shown in Fig.~\ref{fig3}.
Both the temperatures  $T_{\mbox{\scriptsize max}}$ and $T_{\mbox{\scriptsize jet}}$ are very low, around $10^4$~K.
In the following, we make quantitative analysis of how this happens.

From equations (6) and (7), we get 
$$
	\frac{\partial T}{\partial t} \sim \frac{2(\gamma-1)\eta}{\rho \beta_0}j^2
		-C_2\rho\Lambda(T),  \ \ \eqno (12)
$$
where the contribution of the Joule dissipation and radiative losses are 
considered. For simplicity, the first term on the right-hand side of 
Eq.~(12) is denoted by $E_1$, and the second term by $E_2$.
Because both the temperature and its gradient are very small at low 
temperature regions, the heat conduction term ($\sim 10^{-5}$) is neglected 
here. Note that the Joule dissipation and radiative losses play opposite 
roles in the evolution of temperature, one heats the plasma and the other 
cools it. From equation (12), we find that $E_1$ is inversely proportional 
to the density, whereas $E_2$ is proportional to the density. In the 
diffusion region where the density is small, $E_1$ is $\sim 80$, and 
$E_2$ is $\sim 50$. The net effect is that the plasma in the diffusion 
region is heated. However, the heated plasma is pulled out of the diffusion 
region and the cold plasma outside the diffusion region is pushed in, so 
that the plasma in the diffusion region cools. Outside the diffusion region, 
the plasma density along the current sheet increases by several times with 
respect to the initial density (see Fig.~\ref{fig4}), so that $E_2$ increases to 
$\sim 200$ at maximum, resulting in a fast temperature decrease.
{If the plasma temperature is
below $10^4$~K, radiative losses are reduced to zero, so that the relative changes in the plasma
temperature is flat along the outflow region (located along the ``high'' temperature ridge on
the right-hand side of the flux emergence region seen in the bottom right panel of Fig.~\ref{fig3}).
Furthermore, outside the diffusion region, the outflow plasma will expand, which will cool the
plasma. The contribution of the plasma expansion to plasma cooling is
 $(\gamma-1)T\bigtriangledown\cdot{\mathbf v} \sim 40$.
In the diffusion region, the plasma expansion will also cool the plasma there, however
its contribution is very small, only about 4.
A large plasma cooling region is found around $x=2-4$ almost reaching $y=2$ 
(see Fig.~\ref{fig3}), located in the inflow region. This is because of
the plasma expansion in the inflow region, of which the contribution to plasma cooling reaches
40. Even though the contributions of plasma expansion to plasma cooling are of almost the
same order in the inflow region and the outflow region, the plasma temperature decreases
more strongly in the former than in the latter. This is caused by the fact that there is no hot
plasma flowing into the inflow region, whereas hot plasma heated in the diffusion region is
expelled out into the outflow region.} 

It seems that small-scale jets can not originate in a physical environment 
like Exp.~A1 where high mass density and weak magnetic strength are presented, 
as both the velocity and the temperature of the outflow jets are very low.
What will be the result  
if the physical environment with the same magnetic strength but a lower mass 
density is considered? The analysis shows that, if the density is lower, more 
plasma can be heated (because both the radiative losses and the ratio of 
the radiative losses to Joule dissipation is proportional to $\rho^2$).

\begin{figure}[htbp]
\begin{center}
\resizebox{\hsize}{!}{\includegraphics[width=17cm]{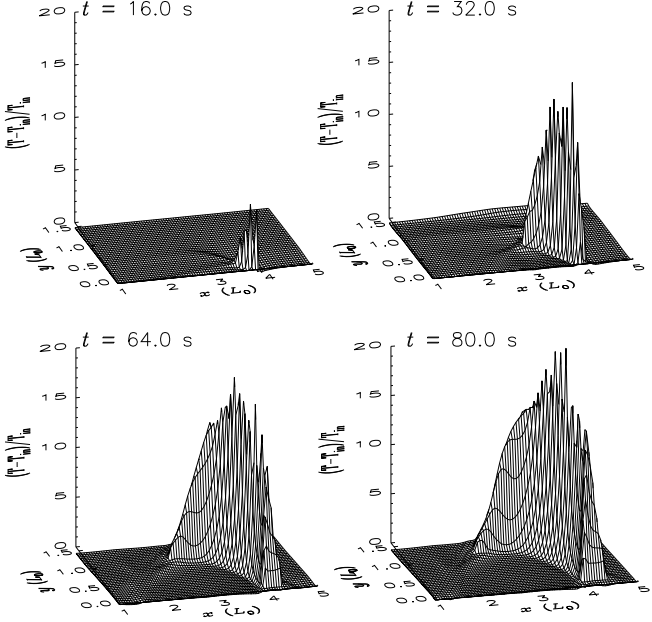}}
\caption{Relative changes in the temperature, 
$(T-T_{\mbox{\scriptsize in}})/T_{\mbox{\scriptsize in}}$ for Exp.~A2, 
with respect to the initial temperature ($T_{\mbox{\scriptsize in}}$),
are shown at four times in the region $1 \le x \le 5, 0 \le y \le 1.6$.}
\label{fig5}
\end{center}
\end{figure}

Fig.~\ref{fig5} shows the evolution of the temperature for the case of Exp.~A2 
where $\beta_0=0.33$ and $\rho_{b0}=0.1$ (this corresponds to
magnetic field strength $B=6$~G, electron density $N_e=2\times 10^{10}$~cm$^{-3}$). 
Significant relative changes in 
temperature are clearly seen in this case and the plasma in the diffusion 
region is heated to the temperature of $T_{\mbox{\scriptsize max}}=2.8\times 
10^5$ K. The maximum velocity of the outflow jet caused by magnetic 
reconnection and its temperature are also analysed for Exp.~A2. We find 
that $v_{\mbox{\scriptsize jet}}$ reaches up to 85 {\rm km s$^{-1}$} and 
$T_{\mbox{\scriptsize jet}}$ is about $2\times 10^5$ K. Both are 
larger than that obtained in Exp.~A1. The direction of $v_{\mbox{\scriptsize 
jet}}$ is $32^\circ$ with respect to the line of sight.

\subsection{Exp.~B ($\beta_0=0.033$)}

\begin{figure}[htbp]
\begin{center}
\resizebox{\hsize}{!}{\includegraphics[width=17cm]{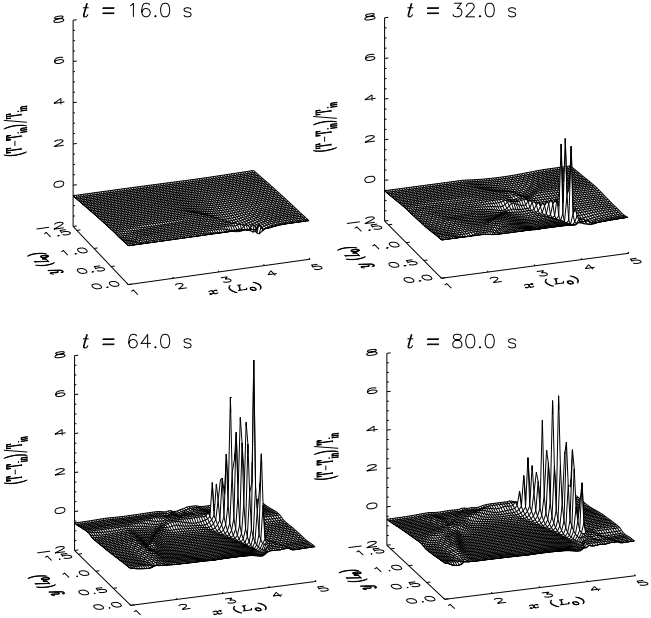}}
\caption{Relative changes in the temperature, 
$(T-T_{\mbox{\scriptsize in}})/T_{\mbox{\scriptsize in}}$ for Exp.~B1, 
with respect to the initial temperature ($T_{\mbox{\scriptsize in}}$),
are shown at four times in the region $1 \le x \le 5, 0 \le y \le 1.6$.}
\label{fig6}
\end{center}
\end{figure}

\begin{figure}[htbp]
\begin{center}
\resizebox{\hsize}{!}{\includegraphics[width=17cm]{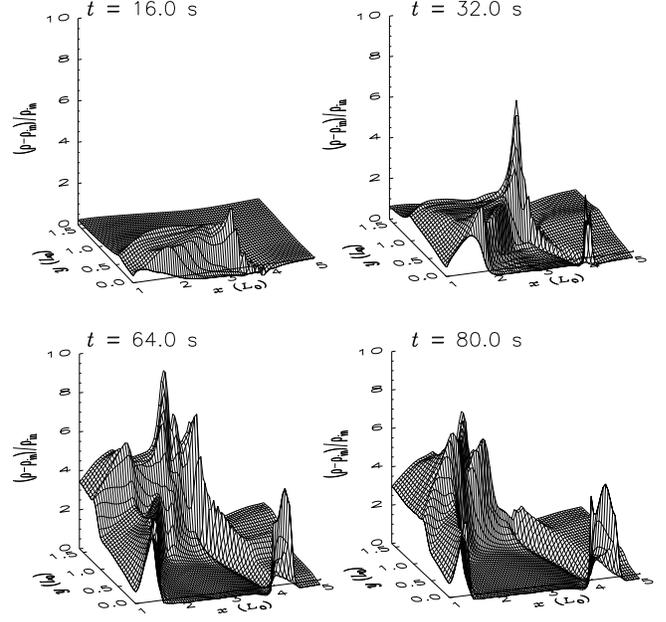}}
\caption{Relative changes in mass density,$(\rho-\rho_{\mbox{\scriptsize in}})/\rho_{\mbox{\scriptsize in}}$
for Exp.~B1, with respect to the initial density ($\rho_{\mbox{\scriptsize in}}$), 
are shown at the same times and in the same region as shown in Fig.~\ref{fig6}.}
\label{fig7}
\end{center}
\end{figure}

In this section, the magnetic reconnection takes place in a physical 
environment with a strong magnetic field. Fig.~\ref{fig6} shows the evolution of 
the temperatures for the case of Exp.~B1 where $\beta_0=0.033$ and $\rho_{b0}=3$ 
which corresponds to $B=20$~G and $N_e=6\times 10^{11}$~cm$^{-3}$.
This represents a physical environment with a strong magnetic field and a high mass 
density. Fig.~\ref{fig7} shows the relative changes in density.
Peak structures are present in the relative changes in both the temperature 
and density on each rim of the flux emergence region. In the diffusion 
region, $T_{\mbox{\scriptsize max}}$ is  $1.5\times 10^5$~K. For the upflow 
jet, $v_{\mbox{\scriptsize jet}}$ reaches $60$~{\rm km s$^{-1}$}, directed 
at $21^\circ$ with respect to the line of sight, and 
$T_{\mbox{\scriptsize jet}}$ is about $8\times 10^4$~K. Comparing the 
results with those in Exp.~A1 where we use the same mass density at the 
bottom but with a weaker magnetic field strength, we find that the plasma 
is heated to higher temperature for the case of Exp.~B1 compared to Exp.~A1.
This is because Joule dissipation becomes stronger as the magnetic field 
strength increases. In the case of Exp.~B1, $E_1\sim 800$, in the diffusion 
region, whereas $E_2\sim 500$. Even though $E_2$ increases as a result of 
the increase of $\Lambda(T)$, the net effect is that more energy is available 
to heat the plasma in the diffusion region. The relative changes in the 
temperature are manifested as a maximum in the diffusion region and 
decrease monotonously in either direction along the current concentration.
The maximum velocity of the outflow jets is larger in Exp.~B1 than in Exp.~A1. 
This is because as the ratio of the magnetic pressure to gas pressure increases, 
it leads to higher acceleration of the outflow plasma as the magnetic field 
strength increases.

\begin{figure}[htbp]
\begin{center}
\resizebox{\hsize}{!}{\includegraphics[width=17cm]{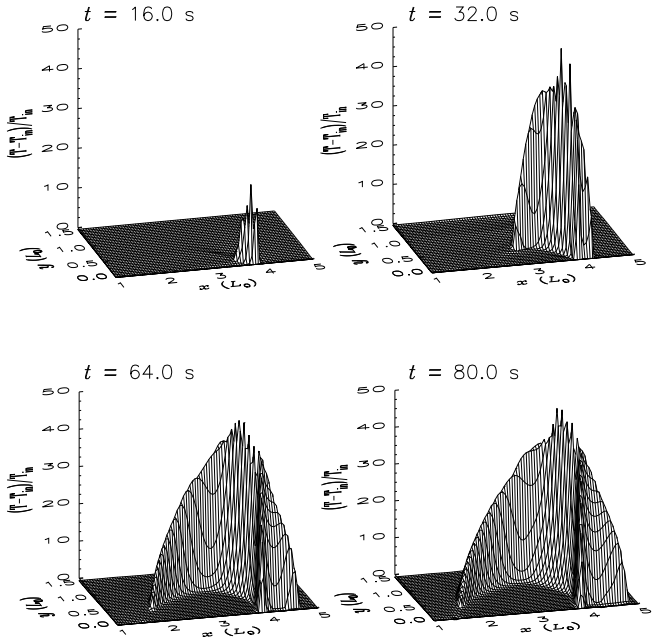}}
\caption{Relative changes in the temperature, 
$(T-T_{\mbox{\scriptsize in}})/T_{\mbox{\scriptsize in}}$ for Exp.~B2, 
with respect to the initial temperature ($T_{\mbox{\scriptsize in}}$),
are shown at four times in the region $1 \le x \le 5, 0 \le y \le 1.6$.}
\label{fig8}
\end{center}
\end{figure}

Fig.~\ref{fig8} shows the results for the case of Exp.~B2, where $\beta_0=0.033$ and $\rho_{b0}=0.1$
(i.e. $B=20$~G and $N_e=2\times 10^{10}$~cm$^{-3}$).
In this case, the plasma in the diffusion region is heated to 
$T_{\mbox{\scriptsize max}}=6.4\times 10^5$~K. The maximum velocity of the 
outflow jets, $v_{\mbox{\scriptsize jet}}$, reaches up to 
$150$~{\rm km s$^{-1}$}, and is 130~{\rm km s$^{-1}$} in the line-of-sight. 
The temperature at the location of $v_{\mbox{\scriptsize jet}}$ is 
$6\times 10^5$~K. Note that, for the case with low mass density, 
the heat conduction term will become more and more important as an energy 
loss in the diffusion region compared to  the radiative loss term, when the 
temperature of the plasma increases to a higher value. 
All the results for both Exp.~A and B are summarized in Table~\ref{table1}.

\begin{table}[htbp]
\caption{Summary of the results for both Exp.~A and B. $v_y$ denotes the line-of-sight component of $v_{\mbox{\scriptsize jet}}$.}
\label{table1}
\centering
\begin{tabular}{c|cccc}
\hline
Quantity & Exp. A1 & Exp. A2	& Exp. B1 & Exp. B2 \\
\hline
$T_{\mbox{\scriptsize max}}$ (K) & $\sim$$1\times 10^4$ & $2.8\times 10^5$ & $1.5\times 10^5$ & $6.4\times 10^5$ \\

$T_{\mbox{\scriptsize jet}}$ (K) & $\sim$$1\times 10^4$ & $2\times 10^5$ & $8\times 10^4$ & $6\times 10^5$ \\

$v_{\mbox{\scriptsize jet}}$ ({\rm km s$^{-1}$}) & 	40		&	85		&	60	& 150 \\

$v_y$ ({\rm km s$^{-1}$})  &	32		&	72	&	56	&	130 \\ 
\hline
\end{tabular}
\end{table}

\section{Discussions} \label{sec::dis}

In the present paper, we investigate magnetic reconnection in the
solar chromosphere, looking for an answer to whether jet-like events can
originate at lower temperatures. We consider magnetic reconnection triggered 
by newly emerging magnetic flux. Several physical environments representing the 
chromosphere with different magnetic field strengths and densities are discussed.
Our results quantitatively show that the temperatures and velocities of the bi-directional jets 
caused by magnetic reconnection are strongly dependent on the physical conditions 
of the atmosphere, such as magnetic field strength and mass density. In the case 
of where the magnetic field strength is high (Exp.~B2) and the density is low, we 
obtain the highest temperatures and the largest velocities among all our simulations.
The plasma in the diffusion region is heated up to $6.4\times 10^5$~K, and the 
maximum velocity of the outflow jets reaches $150$ {\rm km s$^{-1}$}. 
However, in the case of a low magnetic field strength (Exp.~A1) and high density, 
the temperature and the velocities of the jets  are the lowest of all the cases. The 
temperature is about $10^4$~K and the maximum velocity of the jets is only 40 
{\rm km s$^{-1}$}. 
Either an increase in the magnetic field strength or a decrease of the density will raise the 
ratio of the Joule dissipation to the radiative losses. That will heighten the 
heating effect compared to cooling and will lead to a high increase in the 
temperature of the plasma in the diffusion region. That will also raise the ratio 
of the magnetic pressure to the gas pressure, resulting in enhanced 
acceleration of the plasma ejected outward from the diffusion region (which will lead to 
the appearance of high velocity jets). 

{For the two cases, Exp.~A and B,
the amounts of the emerging flux are not the same,
which implies that the stronger the magnetic field strength the larger the amount
of newly emerging magnetic flux.
We can also obtain outflows with high velocity and high temperature
even if a smaller amount of new flux, e.g. the same as that in case A, is emerged in case B.
Moreover, in case B where a larger amount of new flux emerges, 
the temperature and velocity of the outflow jets reach their maximum at about $t=40$~s, 
not at the end of the flux emergence process.
Therefore, the difference in the amount of emerging flux between case A and
B is not very important in determining different results in the two cases presented.
Instead, the magnetic field strength of the initial background magnetic field and 
the initial density structure plays a more significant role.
If we do all the experiments with constant flux emergence, 
the amount of the newly emerging flux will be very small in case B,
compared with that of the background field.
So that the height of the emerging arcade is lower in case B than in case
A at any given time, with the result that reconnection occurs at different heights in the two cases.}

{The flow velocities derived here can be compared with Roussev \etal\ (\cite{roussev01b, roussev01c}) 
who found blue-shifts of the order of 100 \kms 
under the physical environment of electron density $3\times10^{10}$~cm$^{-3}$ and magnetic field 8~G. 
These authors also found that the choice of initial state and the consideration of
non-equilibrium ionization was crucial to their modelling.
Even though various initial physical environments were explored in their work, the plasma
$\beta$ on the current sheet was the same in all the cases.
Both the plasma $\beta$ and the thermal energy 
were assumed to be uniform in the $y-$direction with the reconnection happening in 
the transition region. 
The maximum jet velocities they obtained under different physical environments
did differ at the begining of the experiments, however, they approached to
almost the same value as time went on 
(see top-left panel in Fig.~9 in Roussev \etal\ (\cite{roussev01b})).
In our model, the maximum jet velocities are different for all the cases.
The highest velocity jets are obtained under the environment with almost the same density
as in Roussev \etal\ (\cite{roussev01b}), but
stronger magnetic field (20~G). When the reconnection occurs in the chromosphere, 
high velocity jets-like features are possible if a larger field strength is used.
Besides, we do not consider transient ionization.

In the flux emergence studies of Yokoyama \& Shibata (\cite{yokoy95, yokoy96}) and
Moreno-Insertis \etal\ (\cite{more08}), reconnection occurs at coronal hights
where the plasma temperature is high and the electron density is low ($10^{10}$~cm$^{-3}$),
even though new flux emerges from beneath the photosphere.
Therefore, it is easier to obtain jets with high velocities and high temperatures.
In the Isobe \etal\ (\cite{isobe08}) model, 
magnetic reconnection between newly formed horizontial fields 
and background vertical fields took place at chromospheric heights.
The chromospheric electron density is much higher ($>10^{12}$~cm$^{-3}$) than that in our model,
so that the reconnection jets velocity they obtained is very slow, about 30~\kms.
In the Murray \etal\ (\cite{murray09}) model, 
oscillatory reconnection is driven by the global imbalance of the forces between the 
neighbouring flux systems. At each subsequent reconnection reversal, the system is closer to equilibrium.
Once an equilibrium state is reached, all reconnection ceases,
therefore, the maximum temperature and maximum velocity (50~\kms) achieved in the outflow jets 
decreases in each subsequent reconnection phase.
All the flux emergence studies mentioned above did not include heat conduction and radiative effects.
The latter will reduce both the temperature and the velocity of the outflow jets.

In comparison to Litvinenko \& Chae (\cite{litvi09}) model where radiative losses were assumed to be of 
less importance, they become significantly important
in our model when the density and temperature are large.
Their study was based on the assumption that the inflowing magnetic energy is completely converted
into thermal and kinetic energies of the outflowing plasma.
Therefore, the temperature and speed of the outflow jets are larger in their model
than those in ours at the same height for the same parameter.
}

By using our model we would like to give a possible explanation as to why no 
brightening is obtained at zero velocity of the emission spectral lines in the study 
by Innes \& T\'oth (\cite{innes99}). In their study, the magnetic field strength and mass density 
in the chromosphere was about 6~G and $2.8\times 10^{-10}$ {kg m$^{-3}$}, respectively, 
which corresponds  to $\beta_0=1.3$ and $\rho_{b0}=0.8$ in our model. Our calculations  
show that in this case the maximum temperature in the diffusion region is less 
than $5\times 10^4$~K and therefore, no brightening could appear at the zero velocity 
position of the spectral line.

The radiative loss model that we use at the base is the optical-thin approximation.
McClymont \& Canfield (\cite{mcc83}) considered optical-depth effects and proposed
an optical-thick model for the radiative loss function.
By comparing the two models, 
we find that the optical-depth effects become important primarily for the
plasma at temperatures below $3\times 10^4$~K, where the contribution of hydgron is dominant.
This means, in our model, that the radiative cooling effect is over-evaluated for the cool plasma 
at temperatures below $3\times 10^4$~K. The temperature of the cool plasma
should be raised if we take into account the optical-depth effects in our model.
However, even if we consider optical-depth effects,
the corrected temperature of the cool plasma will not exceed $3\times 10^4$~K.
Otherwise, radiative loss will become too strong and the plasma will start to cool again.
Considering that the jet-like events discussed here correspond to hot jets at transition-region temperatures,
optical depth effects will have less influnce.

{As mentioned in Section 2, background heating is not considered in our model, 
therefore the initial state is not in thermal-equilibrium.
We take the case of Exp A1, shown in Fig.~\ref{fig2}, as an example to examine
the change of the background.
If no emergence takes place, the plasma temperature at the top boundary
will decrease by 27\% by the end of 80~s.
Because of background cooling, downward flow will occur, which
reaches about 40~\kms at maximum during 0--80~s.
The high velocity region is mainly located at high altitude (transition
region or above), whereas the downward flow around the
diffusion region is slower (less than 10~km~s$^{-1}$).}

{Priest \& Forbes (\cite{priest86}) studied mechanisms of magnetic reconnection analytically.
They introduced a new parameter ($b$) and produced a series of analytical solutions. 
They found that the type of reconnection regime
and the rate of reconnection depends sensitively on the parameter $b$
which characterises the inflow conditions.
The Petschek ($b=0$) solution is just one particular member, 
corresponding to a weak fast-model plasma expansion.
However, in our model, there is a fast plasma expansion across the inflow region,
and the diffusion region lengthens as reconnection occurs.
Therefore, the reconnection event here resembles more the fast-mode expansion solutions
with $b>0$ in Priest \& Forbes (\cite{priest86}).

In this study, reconnection occurs on the right-hand side of the emerging flux arcade. If
reconnection occurs on the left-hand side of the emerging flux arcade, e.g. the polarity of the
emerging flux changes, the magnetic field strength on the left-hand side will decrease faster
as reconnection goes on. This is because symmetric boundary conditions are used on the
left, whereas open boundary is used on the right. 
As magnetic field strength decreases, magnetic reconnection
will become weaker. Despite of this, there will not be significant differences in the
results if reconnection occurs on the left-hand side of the emerging flux arcade (because the
jets reach their maximum velocity faster than the decrease of the magnetic field strength). The
magnetic field strength of the initial field on the left-hand side of the flux emergence region is
still very strong, when the jets velocity reach their maximum (for instance, at around 40~s in
case B). If investigating long-term evolution, a larger computational domain, compared with
the flux emergence region, is necessary to weaken the influence of boundary conditions.
}

In the present paper, we mainly concentrate on the hot component of the outflow jets 
caused by magnetic reconnection, and obtain jets with a high temperature and a high 
velocity under certain circumstances. We also produce cool jets in the temperature 
range of $(2\pm 0.2) \times 10^4$~K. Taking the case of Exp.~B2 as an example.
The line-of-sight velocity of the cool upflow jets reaches 25 {\rm km~s$^{-1}$}.
This shows one possibility that some cool jets, e.g. spicules, and hot jets may 
originate from the same phenomenon, chromospheric magnetic reconnection, and they 
may just depict different parts and/or different stages of the same event. 
In the case of weak magnetic field strength but high mass
density, the cool jets disappear, which shows that the origination of spicules may also have some
correlations with the physical environment. We relegate 
further study of the relationship between spicules, jet-like events, and magnetic 
reconnection to future work.

\begin{acknowledgements}

Research at Armagh Observatory is grant-aided by the N.~Ireland Dept. of 
Culture, Arts and Leisure (DCAL). This work was supported via grant ST/F001843/1
from the UK Science and Technology Facilities Council.
MM and JGD thank the International Space Science Institute, Bern for
the support of the team ``Small-scale transient phenomena  and their
contribution to coronal heating''.

\end{acknowledgements}

%\bibliographystyle{aa}
%\bibliography{EE_formation}
\end{document}